\documentclass[reprint,preprintnumbers, amsmath,amssymb, aps]{revtex4-2}

\usepackage{graphicx}
\usepackage{dcolumn}
\usepackage{bm}
\usepackage{changes}
\usepackage{color}
\usepackage{float}
\usepackage{soul}

\begin{document}

\title{Dark Superabsorbers with Dirac-delta-like superdirective radiation}
\author{Jeng Yi Lee}
\affiliation{Department of Opto-Electronic Engineering, National Dong Hwa University, Hualien 974301, Taiwan}

\author{Irving Rond\'on}
\affiliation{School of Computational Sciences, Korea Institute for Advanced Study, Seoul 0245, Republic Of Korea}

\author{Andrey E. Miroshnichenko}
\affiliation{School of Engineering and Information Technology, University of New South Wales, Canberra, Australian Capital Territory 2600, Australia}

\author{Pai-Yen Chen}
\affiliation{Department of Electrical and Computer Engineering, University of Illinois 
 at Chicago,Illinois 60661, USA}

\date{\today}

\begin{abstract} 
We theoretically and numerically reveal that under a given level of extinction  cross section and with definite angular momentum channels dominant, there exists a physical limitation for absorption cross section being maximum  and scattering cross section being minimum.
In addition, any scattering systems operated at this condition would be accompanied by a needle Dirac-delta-like far-field radiation pattern, reducing to perturb the background field except in the forward direction.
We therefore refer to this outcome as dark superabsorbers.
Moreover, by considering the mathematical Gibbs phenomenon, we find that a completely equivalent Dirac-delta far-field radiation is excluded even we could properly design the scatterers operated at such conditions.
We believe this finding has potential applications in design of dark energy harvesting, lower-visibility receivers, superdirective light-matter interaction, and Fresnel diffractive imaging.
\end{abstract}
\pacs{ }

\maketitle

Enhancing light harvesting for any scattering systems is closely related to practical application in  photovoltaics engineering, photodetectors,  photothermal therapy to name a few \cite{energy1,energy2,energy3,energy4,energy5,energy6,energy7,energy8}.
However, scattering of any objects with deep subwavelength scale would suffer from single mode limit, that can be overcame by employing multi-layered structures to induce multipolar resonant modes with spectrally overlapped \cite{perfect1,perfect2,perfect3,perfect4}.
The mechanism is similar to raise degenerate surface wave modes, that can be closely related to the whispering
gallery condition, as already demonstrated in superscattering \cite{scs1,scs2,scs3,scs4}.
In addition, another different mechanism to have superscattering events can get help from constructive interference between one radiation channel at bound state in the continuum and another channel at resonance \cite{bic1}.
We note that with a consideration of energy conservation, there has an ultimate bound on partial absorption cross section (ABS) at each scattering channel \cite{bound1,bound2,bound3}.
On  the other hand, we notice that enhancing ABS would pay the price on the accumulation of extinction cross section (EXT).
 EXT is related to fundamental optical theorem, that can measure the amount of forward scattering amplitude in parallel with the incident wave \cite{book3,kerker2}.
 One example to discuss the relation of EXT and forward scattering amplitude is the Kerker effect, in which combining $\pi$- or $0$- phase shift for electric and magnetic dipolars with equal strength would lead to have optimized backward or forward scattering, respectively \cite{kerker1,kerker2,kerker3,kerker4,kerker5,kerker6,kerker7,kerker8,kerker9}.
On the other hand, based on Lorentz reciprocity, any scattering objects made of reciprocal materials would preserve EXT at two opposite illuminating directions \cite{extsymmetry}. 
Nevertheless such reciprocity can be broken by exploiting  optically nonlinear materials \cite{reciprocity1}.

In this work, we would indicate that under a constant level of EXT, there could have a strategy to design the scatterers with not only ABS enhanced as well as scattering cross section (SCS) reduced, 
but also an unusual Dirac-delta-like far-field radiation.
We  referred to the outcome as dark superabsorbers.
Our scheme is firstly by means of normalized cross section diagram, stemming from optimizing an energy function involving SCS and ABS, as shown in Fig. 1 (b) \cite{bound3}.
Accordingly, we can indicate generic power distribution for ABS, SCS, and EXT for any scattering systems, irrespective of system parameters.
Next, with a consideration of a constant level of EXT and definite angular momentum channels dominant, we can point out the maximum ABS and the minimum SCS.  
Interestingly, at this condition, we find that the description of the far-field radiation pattern is mathematically equivalent to a discrete Fourier series of the  Dirac-delta function.
Moreover, the corresponding null-to-null bandwidth of the main lobe would be gradually reduced by more angular momentum channels properly excited, resulting in perturbing the background field except in the forward direction.
However, by considering the Gibbs phenomenon of Fourier theory, the ideal (perfect) Dirac-delta radiation is excluded.
Instead, it would have many ripple-like sidelobes around the main lobe.
We also numerically find that the ratio of the total radiation power of these sidelobes over that of the main lobe approaches a constant as more angular momentum channels are properly excited.  
We  believe our finding would benefit possible applications on dark energy harvesting, lower-visibility receivers, directive light-matter interaction, and Fresnel diffractive imaging.

 \begin{figure*}[ht]
\centering
\includegraphics[width=1\textwidth]{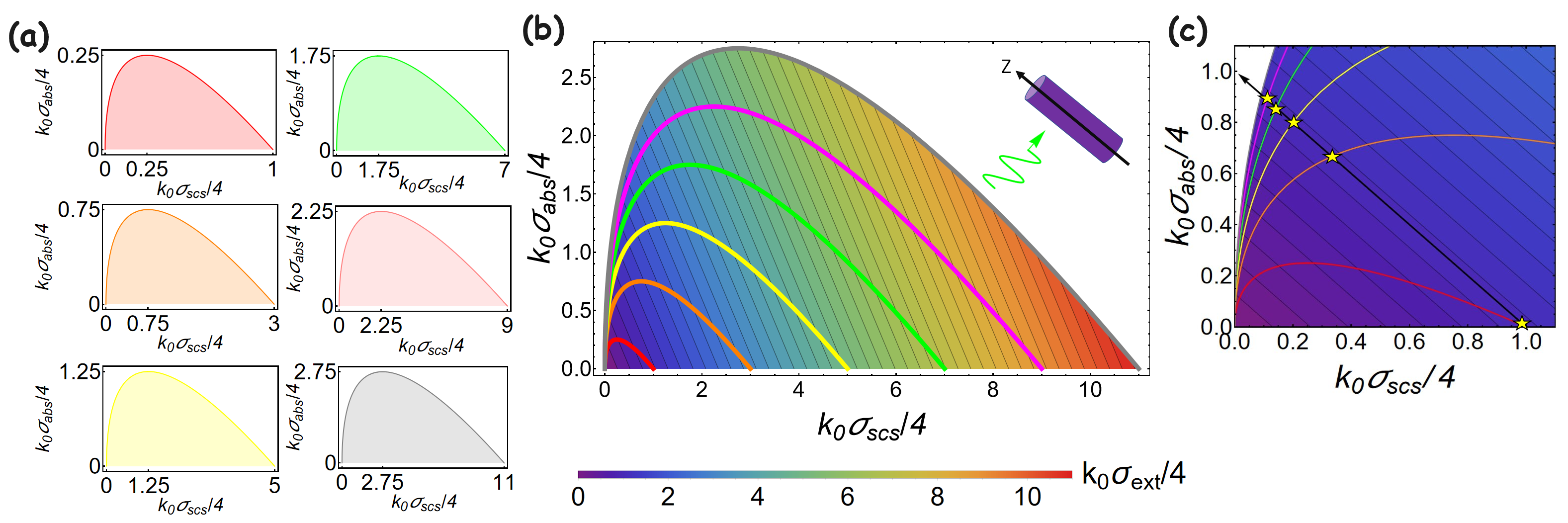}
\caption{Physical boundary and domain of normalized SCS ($k_0\sigma_{scs}/4$) and normalized ABS ($k_0\sigma_{abs}/4$) for $N=0$,  $N=1$, $N=2$, $N=3$, $N=4$, $N=5$ dominant in (a).
Integrating these results and considering EXT ($\sigma_{ext}=\sigma_{abs}+\sigma_{scs}$), the contourplot of normalized EXT ($k_0\sigma_{ext}/4$) is provided in (b). Under a constant level of $k_0\sigma_{ext}/4=1$, by more N channels involved, the levels of normalized SCS and normalized ABS would be further reduced and increased, respectively, as highlighted by yellow stars in (c). }
\end{figure*}

We consider a cylindrical scatterer illuminated by a normal incidence of an electromagnetic plane wave with linear polarization.
Below, we discuss the $\textbf{p}$-polarized excitation, in which the exciting magnetic field oscillates only along the z-direction, as shown in an inert of Fig. 1 (b).
The formulas of ABS  and SCS for arbitrary cylindrical scatterers are $\sigma_{abs}=\sum_{n=-N}^{N}\sigma_{abs,n}=-4/k_0\sum_{n=-N}^{N}[Re[a_n]+\vert a_n\vert^2]$ and $\sigma_{scs}=\sum_{n=-N}^{N}\sigma_{scs,n}=4/k_0\sum_{n=-N}^{N}\vert a_n\vert^2$, respectively \cite{book1,book2}. 
Here $\sigma_{abs,n}$ and $\sigma_{scs,n}$ are the partial ABS and SCS, $a_n$ is complex scattering coefficient for n-th angular momentum channel, $k_0$ is  background wavenumber defined by $k_0=2\pi/\lambda_0$, and $\lambda_0$ is operating wavelength. 
We mark that the notation of $N$ refers to the terminate term of the dominant cylindrical wave scattering, while $n=0,1,2,..$ corresponds to  magnetic dipole, electric dipole, and electric quadrupole, respectively. 
On the other hand, EXT is the summation of ABS and SCS, i.e., $\sigma_{ext}=\sigma_{abs}+\sigma_{scs}$, also linking the relation of forward scattering amplitudes \cite{book3,kerker2}.
Thus, any passive scatterers with non-zeros ABS and SCS would accumulate EXT, causing a shadow.
Moreover, due to the cylindrical symmetry at normal incidence, there has a degeneracy for $a_{n}=a_{-n}$.

As generic scattering systems made of passive materials embedded, from energy conservation, it ensures partial ABS subject to $\sigma_{abs,n}\geq 0$. 
Consequently, we can obtain a phasor condition for complex scattering coefficients established at each angular momentum channels, i.e., $\vert a_n\vert\leq 1$ and $Arg[a_n]\in [0,2\pi)$ \cite{bound2,bound3}.
Accordingly, we can point out the maximum partial ABS for each angular momentum channels operated at $a_n=1/2$ and $Arg[a_n]=\pi$, corresponding to the eigenvalue zero of the scattering matrix, i.e., the  coherent perfect absorption condition \cite{perfectabs1,perfectabs2,perfectabs3,perfectabs4}.
In addition, we also find the condition of the maximum partial SCS, when the scattering systems are operated at $a_n=1$ and $Arg[a_n]=\pi$.

The comprehension of power distribution and its physical limitation for any scattering objects with any channels excited is desirable, which would benefit designs of functional devices.
In our work of \cite{bound3}, we employ the concept of optimization to address such doubt.
We now consider a constant level of $\sigma_{abs}$ at definite N angular momentum channels excited, then we seek to evaluate the extreme of $\sigma_{scs}$.
 We thus define an energy function $L$ involving SCS and ABS as follows, i.e., 
 $L(a_{-N},..,0,..a_{N})=\sigma_{scs}+\lambda \sigma_{abs}$,
 where $\lambda$ is a Lagrange undetermined multiplier.
 To have the extreme, the energy function needs to simultaneously satisfy $\partial L/\partial \vert a_i\vert=0$ and $\partial L/\partial \theta_i=0$ for $i=-N\sim N$. Here we express the complex scattering coefficient as $a_i=\vert a_i\vert e^{i\theta_i}$ where $\theta_i$ is the argument.
 As a result, we find that under a constant  $\sigma_{abs}$ and with N angular momentum channels involved, there has two extremes of $\sigma_{sc}$ with one being minimum and another being maximum.
We indicate such extreme conditions depicted in the solid lines of Figs. 1 (a)-(b).
We notice that any scattering systems operated at such conditions would have $\vert a_i\vert =constant$ and $\theta_i=\pi$ for $i=-N\sim N$.
With these outcomes, we depict the corresponding physical boundary and domain for total normalized SCS defined as $k_0\sigma_{scs}/4$ and normalized ABS defined as $k_0\sigma_{abs}/4$  in Fig. 1 (a),
for $N=0$, $N=1$, $N=2$, $N=3$, $N=4$, and $N=5$ dominant marked by red, orange, yellow, green, pink, and gray colors.
\begin{figure}[ht]
\centering
\includegraphics[width=0.5\textwidth]{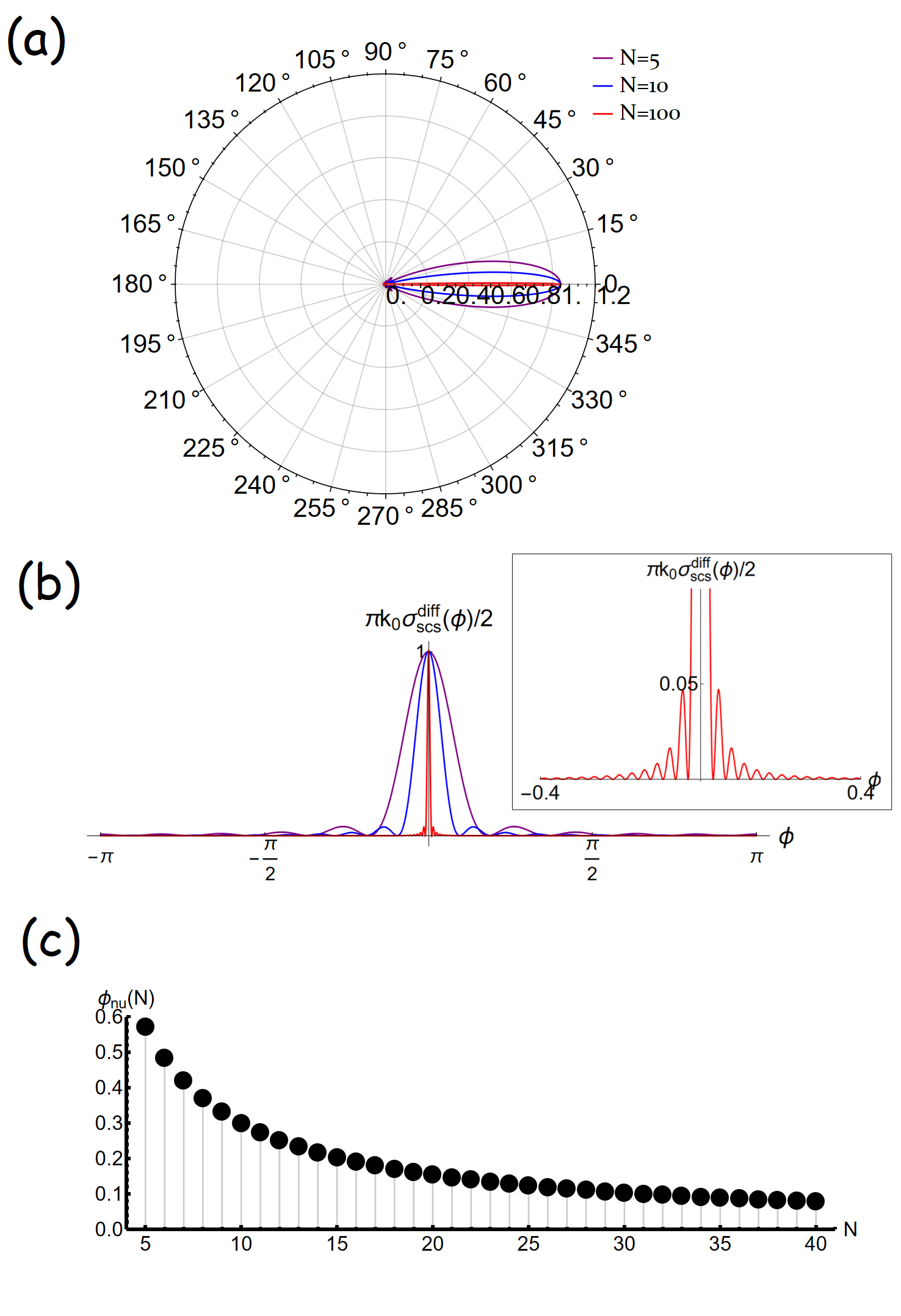}
\caption{By considering $N=[5,10,100]$ and $k_0\sigma_{ext}/4=1$  into Eq.(3), we depict the corresponding far-field radiation in (a) and (b). Here these results are expressed in units of $\pi k_0\sigma_{scs}^{diff}(\phi)/2$.
In the insert of (b), we highlight the ripple-like sidelobes, around the main lobe, corresponding to the Gibbs phenomenon.
In (c), we calculate the null-to-null bandwidth of the main lobe, i.e., $\phi_{nu},$ for $5\leq N \leq 40$. }
\end{figure}

As arbitrary scattering systems with $N=0$, $N=1$, $N=2$, $N=3$, $N=4$, and $N=5$ dominant, the corresponding maximum normalized SCS, i.e., $k_0\sigma_{scs}/4$, would be $1$, $3$, $5$, $7$, $9$, and $11$, respectively, as clearly shown in Fig. 1 (a).
The maximum normalized ABS, i.e., $k_0\sigma_{abs}/4$, for $N=0$, $N=1$, $N=2$, $N=3$, $N=4$, and $N=5$ dominant, would be $1/4$, $3/4$, $5/4$, $7/4$, $9/4$, and $11/4$, respectively, as clearly shown in Fig. 1 (a).
Next, we integrate all results of Fig. 1 (a) into a single normalized cross section diagram in Fig. 1 (b).
In addition, by considering $\sigma_{ext}=\sigma_{abs}+\sigma_{scs}$, we also make the contour plot for the normalized EXT defined as $k_0\sigma_{ext}/4$ in Fig. 1 (b).
With more N angular momentum channels involved, the domain for SCS, ABS, and EXT would be further extended as well.
Moreover, we can see that due to the fact that the domain of larger angular momentum channels surrounds that of lower ones, any scattering systems with more channels involved may behave the same cross sections as by lower channels, exhibiting the degeneracy.
In Fig. 1 (b), we observe that along a constant level of normalized EXT, i.e., $k_0\sigma_{ext}/4=constant$, there enables the gradual decrease of normalized SCS as well as the gradual increase of normalized ABS.

Now, we turn to discuss the corresponding Far-field radiation at the boundary of $N=1$, $N=2$, $N=3$, and so on. 
With a constant normalized EXT and along the N-boundary as in Figs. 1 (a)-(b),
the scattering coefficients would satisfy,
\begin{equation}
a_i=-\frac{k_0\sigma_{ext}}{4(2N+1)},i\in[-N,N].
\end{equation}
The corresponding SCS and ABS would be
\begin{equation}
\begin{cases}
\begin{split}
\sigma_{scs}(N)&=\frac{k_0\sigma_{ext}^2}{4(2N+1)}\\
\sigma_{abs}(N)&=\sigma_{ext}[1-\frac{k_0\sigma_{ext}}{4(2N+1)}],
\end{split}\end{cases}
\end{equation}
 and the corresponding differential SCS is 
\begin{equation}
\sigma_{scs}^{diff}(\phi)=\frac{ k_0\pi\sigma^2_{ext}}{2(2N+1)^2}\vert \frac{1}{2\pi} \sum_{n=-N}^{N}e^{in\phi}\vert^2
\end{equation}
\cite{book2}, where $\phi$ is the azimuth of  cylindrical coordinate system.
Interestingly, we find that the term of $\frac{1}{2\pi} \sum_{n=-N}^{N}e^{in\phi}$ as $N\rightarrow \infty$ is mathematically equivalent to the discrete Fourier series of the Dirac-delta function i.e.,
\begin{equation}
\begin{split}
\delta(\phi)&=\frac{1}{2\pi}+\frac{1}{\pi}\sum_{n=1}^{\infty}\cos[n\theta]=\frac{1}{2\pi}\sum_{n=-\infty}^{\infty}e^{in\phi}
\end{split}
\end{equation}
\cite{book5}.
This result reflects that any scattering systems operated at the N-th boundary  would perform a needle Dirac-delta like far-field radiation pattern.
We note that such Dirac-delta of radiation,  with highly super-directive, had discovered by degenerate resonant channels excited \cite{superdirective}.

\begin{figure*}[ht]
\centering
\includegraphics[width=0.85\textwidth]{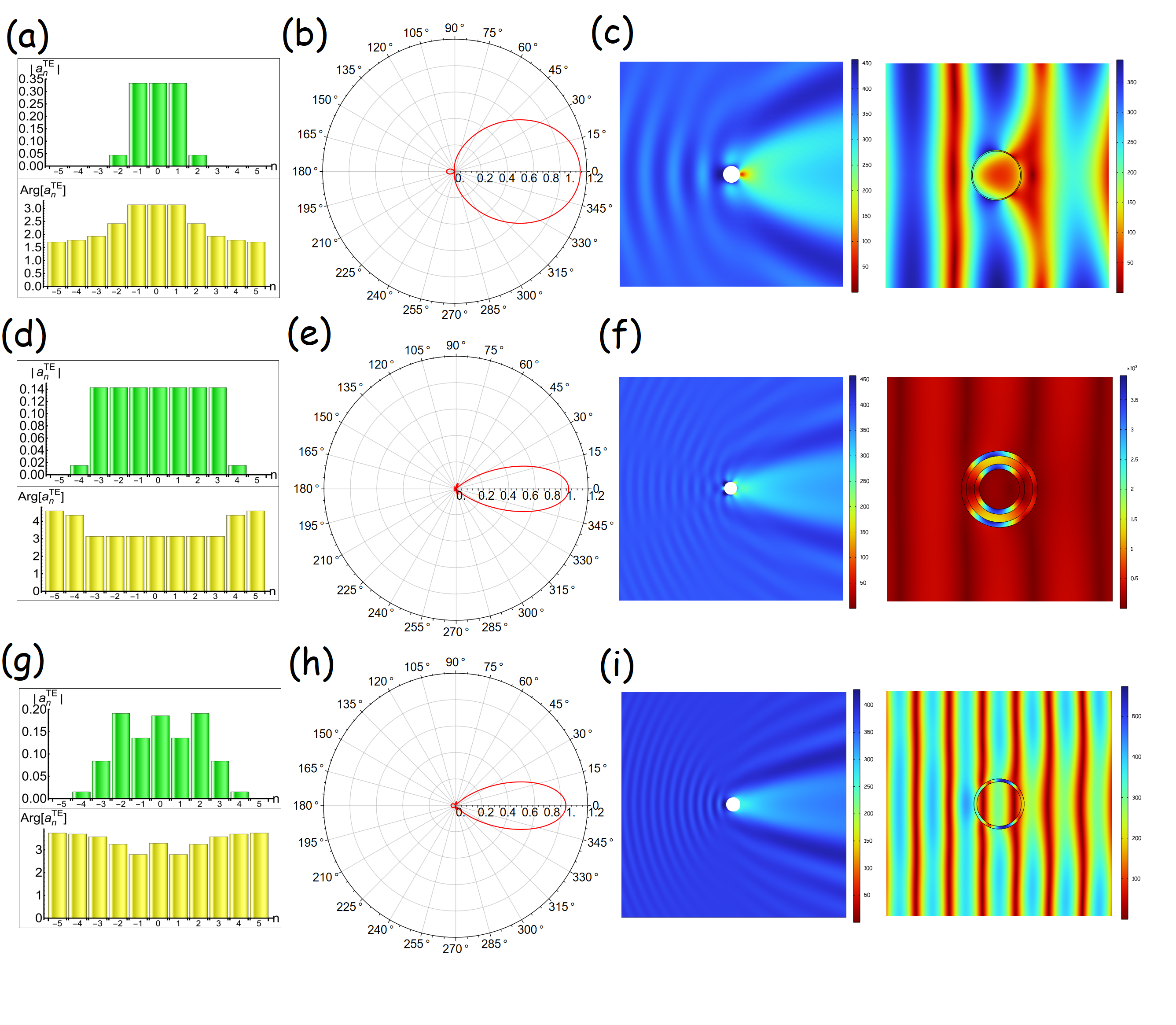}
\caption{By designing a two-layered cylindrical nanowire with the geometry of $r_1=0.142\lambda_0$ and $r_2=0.149\lambda_0$ and the material parameters of $\epsilon_1=-4.7+0.058i$ and $\epsilon_2=51.342+23.986i$, we provide the multipole analysis of $a_n$ in (a). By exploiting COMSOL Multiphysics, the corresponding far-field radiation, far-field and near-instant field distribution for this design are in shown in (b) and (c).
In (d)-(f), there are multipole analysis of $a_n$, far-field radiation,  far-field and near-instant field distribution for another design of a four-layered cylindrical nanowire with the geometry of $r_1=0.151\lambda_0$, $r_2=0.189\lambda_0$, $r_3=0.249\lambda_0$, and $r_4=0.287\lambda_0$ as well as the material parameters of $\epsilon_1=5.149 + 0.561 i$, $\epsilon_2=0.113+0.011i$, $\epsilon_3=0.302+ 0.037i$, and $\epsilon_4=-0.107+0.002i$. 	
Then we design another core-shell cylindrical system by the geometry of $r_1=0.333\lambda_0$ and $r_2=0.378\lambda_0$ and the material parameters of $\epsilon_1=1.157+0.1i$ and $\epsilon_2=0.368+0.4i$. In (g)-(i), we depict the corresponding multipole analysis, far-field radiation,  far-field, and near-instant field distribution.}
\end{figure*}

To demonstrate this result, we choose $k_0\sigma_{ext}/4=1$ followed with a black arrow in Fig. 1 (c).
Then with N   channels involved, the values of ($k_0\sigma_{scs}/4,k_0\sigma_{abs}/4$)  for $N=0$, $N=1$, $N=2$, $N=3$, $N=4$, $N=5$, would correspond to $(1,0)$, $(1/3,2/3)$, $(1/5,4/5)$, $(1/7,6/7)$, $(1/9,8/9)$, $(1/11,10/11)$, respectively, as marked by yellow stars in Fig. 1 (c).
We can see that properly exciting more $N$ channels, there leads to have $\sigma_{scs}\rightarrow 0$ and $\sigma_{abs}\rightarrow \sigma_{ext}$.
Then, we depict the far field radiation distribution in Figs. 2 (a)-(b), by considering $N=[5,10,100]$, respectively.
We can see that except far-field radiation at the forward direction ($\phi=0$) remains fixed attributed to the fundamental optical theorem, the width of the main lobe gets reduced by increasing $N$ channels.
Moreover, the sidelobes get pushed toward the main lobe with respect to N channels, causing an accumulation of ripple around the main lobe, as clearly shown in the insert of Fig. 2 (b).
In mathematics, it corresponds to the Gibbs phenomenon, resulting from the discontinuity of a function \cite{book5}.
We also calculate the null to null bandwidth of the main lobe, i.e., $\sigma_{scs}^{diff}(\phi_{nu})=0$, as shown in Fig. 2(c).
The result indicates that by more angular momentum channels involved, 
$\phi_{nu}$ would get reduced, 
reducing to perturb the background field.

Now, we design such anomalous subwavelength nanowires not only with ABS enhanced as well as SCS reduced, but also with the needle Dirac-delta-like far-field radiation.
We design non-magnetic, passive, and reciprocal nanowires with multi-layered structures.
Then we target  $a_n=-1/3$ at $n=[-1,0,1]$.
As we design such system with $r_1=0.142\lambda_0$ (core) and $r_2=0.149\lambda_0$ (shell), we numerically find that $\epsilon_1=-4.7+0.058i$ in core and $\epsilon_2=51.342+23.986i$ in shell can meet the conditions.
We provide the multipole  analysis of $a_n$ in Fig. 3 (a), consistent with our desired setting.
However, there has a small residue at $a_{2}$ and $a_{-2}$, whose magnitude is quite small compared with $a_n=-1/3$ at $n=[-1,0,1]$.
With COMSOL Multiphysics, we then plot the corresponding far-field radiation pattern of $\sigma_{scs}^{diff}(\phi)$ in Fig. 3 (b), where a small sidelobe is found in the backward direction ($\phi=\pi$).
We  numerically calculate the null-to-null bandwidth of $\phi_{nu}$ being $2.042$.
We also plot the far- and near- field distribution for this two-layered nanowire, where we can observe that a quite large disturbance occurs at the backward and forward directions.
In this case, the normalized SCS ($k_0\sigma_{scs}/4$), normalized ABS ($k_0\sigma_{abs}/4$), and normalized EXT ($k_0\sigma_{ext}/4$) are $0.337$, $0.729$, and $1.066$, respectively.
We mark that this case is superabsorption, since $k_0\sigma_{abs}/4=0.729$ is larger than a single channel limit of $k_0\sigma_{abs}/4=0.25$.

Next, we seek to target the condition of $a_n=-1/7$ at $n=[-3,-2,-1,0,1,2,3]$, belonging to $N=3$ domain.
We find that a four-layered nanowire can fulfil this desired design, with the geometry of $r_1=0.151\lambda_0$, $r_2=0.189\lambda_0$, $r_3=0.249\lambda_0$, and $r_4=0.287\lambda_0$ and the material parameters of $\epsilon_1=5.149 + 0.561 i$, $\epsilon_2=0.113+0.011i$, $\epsilon_3=0.302+ 0.037i$, and $\epsilon_4=-0.107+0.002i$. 	
Then we provide the multipole  analysis of $a_n$ in Fig. 3 (d), meeting our requirement of $a_{n}=-1/7$ from $n=-3$ to $n=3$, although there still has a small residue from $a_{-4}$ and $a_{4}$.
We plot the corresponding far-field radiation pattern in Fig. 3 (e), here the radiation is largely reduced except in the forward.
Compared with the previous design in Figs. 3 (a)-(c), this nanowire system can exhibit more shaper far-field radiation with $\phi_{nu}=0.889$, although they all have the same level of EXT.
We plot the far- and near- field distributions in Fig. 3 (f).
For a comparison, we can observe a more reducing disturbance of the scattering field on the background field.
We notice that this scattering system exhibits  ABS enhanced and SCS reduced. 
Here $k_0\sigma_{scs}/4=0.143$, $k_0\sigma_{abs}/4=0.868$,  and $k_0\sigma_{ext}/4=1.011$.
In the near field distribution, we observe that there has giant electric fields occurred in the shell layers of $r_1\leq r \leq r_2$ and $r_3\leq r\leq r_4$,  up to $3.5\times 10^{3}[V/m]$, with possible application in strengthening light-matter interaction and nonlinear optics.
The extreme field in these shell layers is attributed to the localized surface plasmon excited.

\begin{figure}[ht]
\centering
\includegraphics[width=0.5\textwidth]{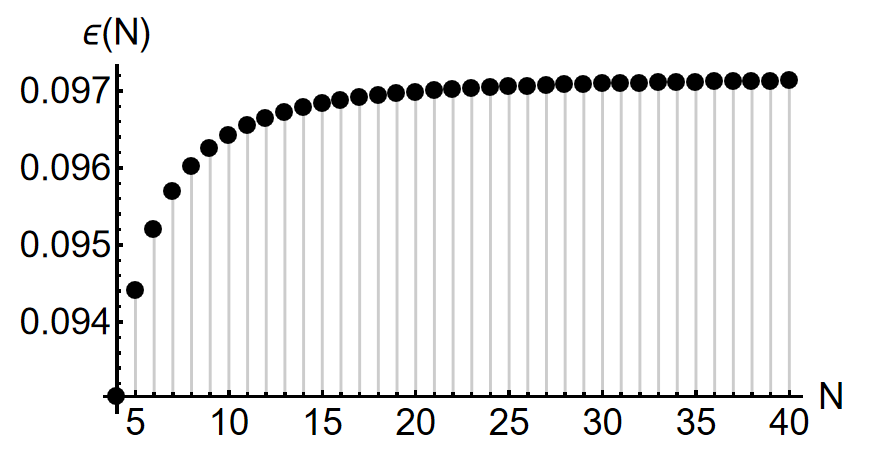}
\caption{A radiation ratio of Eq. (5) with respect to $N$. }
\end{figure}

Now we arrive at two crucial doubts:\\
 (i) Could there exist an ideal (perfect) needle Dirac-delta scattering radiation, if $a_n$ can be properly excited ?\\
(ii) Without following Eq.(1), could any scatterers having near zero of SCS and $\sigma_{abs}\rightarrow\sigma_{ext}$ perform a needle like scattering pattern?\\
To address (i), we draw support from the Gibbs phenomenon of Fourier theory.
Therefore, the formation of the ideal (perfect) needle radiation is excluded, even if one can unlimitedly and properly excite any number of angular momentum channels.
Instead, one can have  a close affinity of Dirac-Delta like far-field radiation with extreme small ripple-like sidelobes around the main lobe.\\
To address (ii), we resort to the theory of Fourier transform theory and Gibbs phenomenon.
Since any dominant $a_n$ following Eq.(1) would have the Dirac-delta-like shape of radiation, any deviation of the Fourier coefficients would deviate to have a needle-like shape.
Therefore, any scatterers even with  largely reducing SCS and $\sigma_{abs}\rightarrow\sigma_{ext}$ can not guarantee to have a needle-like radiation.
Based on our dark superabsorber scheme, we thus calculate a ratio defined as the total radiation from sidelobes over SCS, as follows,
\begin{equation}
\begin{split}
\epsilon(N)&=\frac{\int_{\phi_{nu}}^{2\pi-\phi_{nu}}\sigma_{scs}^{diff}(\phi)d\phi}{\sigma_{scs}(N)}\\
&=\frac{1}{2\pi(2N+1)}\int_{\phi_{nu}}^{2\pi-\phi_{nu}}\vert \sum_{n=-N}^{N}e^{in\phi}  \vert^2 d\phi.
\end{split}
\end{equation}
We provide the numerical results with respect to N in Fig. 4 (a), where we can see that increasing $N$ channels, this ratio would roughly be $0.097$.
We observe that this result is independence of $\sigma_{ext}$.
It reveals that the radiation from these sidelobe becomes a residual source for other scatterers without following our scheme.
As a result, we expect that these scatterers would radiate not only in the forward direction, but also over other directions.
We demonstrate such subwavelength scatterer by using two-layered structure in Figs. 3 (g)-(i) with geometry of $r_1=0.333\lambda_0$ and $r_2=0.378\lambda_0$ and material parameters of $\epsilon_1=1.157+0.1i$ and $\epsilon_2=0.368 + 0.4i$.
In this case, it has $k_0\sigma_{scs}/4=0.159$, $k_0\sigma_{abs}/4=0.84$, and $k_0\sigma_{ext}/4=0.999$ as almost the same values as demonstrated in Figs. 3 (d)-(f).
However, as we provide the multipole analysis of $a_{n}$, in Fig. 3 (g), we find that this case can not fulfill the Eq.(1). 
Then we depict the far-field radiation in Fig. 3 (h), where we can observe a small sidelobe occurred at the backward.
In Fig. 3 (i), we provide the corresponding far- and near- field distribution, showing the significant disturbance to the background field in the backward and forward directions.


The aim to achieve superdirective scattering radiation has implemented in Yagi-Uda-inspired nanoantennas, Huygens-resonantor array, high-order-assisted multipolar resonants to name a few, with benefit of applications in enhanced emission of quantum dots, single photon emission, wireless power delivery, and Raman scattering \cite{antenna1,antenna2,antenna3,antenna4,antenna5,
antenna6,antenna7,antenna8,antenna9,antenna10,
antenna11,antenna12,antenna13,antenna14,antenna15}.
Here we demonstrate a whole new scheme to not only have superdirective needle-like radiation, but also have suppressed sidelobes.
The resultant outcome is to reduce the disturbance on the background fields, except in the forward.
The design of scatterers having  maintaining lower-scattering but without sacrificing absorption had been theoretically and experimentally demonstrated, with practical applications on lower-visibility sensors, Fresnel diffractive imaging, and so on \cite{new1,new2,new3,new4,new5,new6,new7}.
As our demonstration in the case of Figs. 3 (g)-(i), targeting the extreme ABS and SCS for such specific designs is insufficient to meet low-observability.

To our summary, we formulate the conditions for any scattering systems not only  with a near same level of ABS as EXT ($\sigma_{abs}\rightarrow \sigma_{ext}$) but also with the Dirac-delta-like far-field radiation, resulting in a reducing distance to the background field except in the forward direction.
We refer to the results as dark superabsorbers.
Resorting to Gibbs phenomenon of Fourier theory, the perfect single needle-like far-field radiation is excluded, while there are always many ripple sidelobes around the main lobe.
We discuss other possible scattering events having enhanced ABS and reduced SCS.
We argue that our proposed scheme could offer an optimized reduced perturbation to the background field.


\section*{Acknowledgements}

This work was supported by Ministry of Science and Technology, Taiwan (MOST) ($107$-
$2112$-M-$259$-$007$-MY3, $110$-$2112$-M-$259$-$005$- and$111$-$2112$-M-$259$ -$011$ -).

\end{document}